\documentstyle[times]{mn}
\input{epsf}
\newcommand{\bez}{\begin{eqnarray*}}
\newcommand{\eez}{\end{eqnarray*}}
\newcommand{\be}{\begin{equation}}
\newcommand{\ee}{\end{equation}}
\newcommand{\beq}{\begin{eqnarray}}
\newcommand{\eeq}{\end{eqnarray}}
\newcommand{\bc}{\begin{center}}
\newcommand{\ec}{\end{center}}

\newbox\grsign \setbox\grsign=\hbox{$>$} \newdimen\grdimen \grdimen=\ht\grsign
\newbox\simlessbox \newbox\simgreatbox \newbox\simpropbox
\setbox\simgreatbox=\hbox{\raise.5ex\hbox{$>$}\llap
     {\lower.5ex\hbox{$\sim$}}}\ht1=\grdimen\dp1=0pt
\setbox\simlessbox=\hbox{\raise.5ex\hbox{$<$}\llap
     {\lower.5ex\hbox{$\sim$}}}\ht2=\grdimen\dp2=0pt
\setbox\simpropbox=\hbox{\raise.5ex\hbox{$\propto$}\llap
     {\lower.5ex\hbox{$\sim$}}}\ht2=\grdimen\dp2=0pt
\def\simgt{\mathrel{\copy\simgreatbox}}
\def\simlt{\mathrel{\copy\simlessbox}}

\def\bbeta{\mbox{\boldmath $\beta$}}

\def\lcr{l_{\rm cr}}
\def\sT{\sigma_{\rm T}}

\def\RC{R_{\rm C}}
\def\tC{t_{\rm C}}
\def\LE{L_{\rm E}}
\def\dd{{\rm d}}
\def\ctg{{\rm ctg}}
\def\dM{\dot{M}}
\def\dm{\dot{m}}
\def\dmE{\dot{m}_{\rm E}}
\def\dMt{\dot{M}_{\rm tot}}

\def\thi{\theta_\infty}
\def\vpi{\varphi_\infty}
\def\vff{v_{\rm ff}}

\def\tep{t_{\rm ep}}
\def\jK{j_{\rm K}}
\def\lmin{l_*}

\def\rmax{r_{\rm max}}
\def\rcr{r_{\rm cr}}
\def\tT{\tau_{\rm T}}

\begin{document}

\title[Small-scale inviscid accretion discs]
{Small-scale inviscid accretion discs around black holes}

\author[Andrei M. Beloborodov and Andrei F. Illarionov]
{\parbox[]{6.8in}{Andrei M. Beloborodov$^{1,2}$ and Andrei F. 
Illarionov$^{1,2}$}\\
$^1$Stockholm Observatory, SE-133 36 Saltsj\"obaden, Sweden\\
$^2$Astro-Space Center of Lebedev Physical Institute,
  Profsoyuznaya 84/32, 117810 Moscow, Russia}

\date{Accepted, Received}

\maketitle

%This is single spacing
%\baselineskip 12pt
%This is double spacing
%\baselineskip 24pt

\begin{abstract}
Gas falling quasi-spherically onto a black hole forms an 
inner accretion disc if its specific angular momentum $l$ exceeds 
$\lmin\sim r_gc$ where $r_g$ is the Schwarzschild radius. 
The standard disc model assumes $l\gg\lmin$. 
We argue that, in many black-hole sources, 
accretion flows have angular momenta just above the threshold for disc
formation, $l\simgt\lmin$, and assess the accretion mechanism in this regime.
In a range $\lmin<l<\lcr$, a small-scale disc forms in which gas spirals 
fast into the black hole without any help of horizontal viscous stresses. 
Such an `inviscid' disc, however, interacts inelastically with the feeding 
infall. The disc-infall interaction determines the dynamics and luminosity of 
the accretion flow. The inviscid disc radius can be as large as $14 r_g$, and 
the energy release peaks at $2r_g$. 
The disc emits a Comptonized X-ray spectrum with a break at $\sim 100$~keV. 
This accretion regime is likely to take place in wind-fed X-ray binaries and 
is also possible in active galactic nuclei.
\end{abstract}

\begin{keywords}
{accretion, accretion discs --- 
binaries: general ---
black hole physics --- 
radiative mechanisms: thermal ---
X-rays: galaxies ---
X-rays: stars
}  
\end{keywords}

%\footnotetext{$^\star$E-mail: andrei@astro.su.se}

%##########################################################################

\section{Introduction}

Spherical accretion flows onto black holes (BHs) have low radiative 
efficiencies (Shvartsman 1971; Shapiro 1973; but see also Bisnovatyi-Kogan
\& Ruzmaikin 1974; M\'esz\'aros 1975).
If the flow is slightly rotating, the situation changes crucially: 
a small-scale accretion disc forms  
and then the BH `switches on' as a luminous source. 
The condition for disc formation 
reads $l>\lmin$ where $l$ is the specific angular 
momentum of the accreting matter and $\lmin=0.75 r_gc$ (see Section~2.2),
$r_g=2GM/c^2$ being the Schwarzschild radius.
The standard disc model assumes $l\gg \lmin$ (see Pringle 1981 for a review).
High-$l$ discs can form, e.g., in X-ray binaries fed by streams from 
Lagrangian $L_1$ point.

By contrast, in wind-fed X-ray binaries, the accretion flows are 
quasi-spherical with $l\sim l_*$
(see e.g. Frank, King, \& Raine 1992 for a review).
The average $l$ of the accreting wind matter is  
$\bar{l}=\zeta(1/4)\Omega R_a^2$ (Illarionov \& Sunyaev 1975; Shapiro \& 
Lightman 1976) where $R_a=2GM/w^2\sim 10^{11}$~cm is the accretion 
radius, $w\approx 10^8$~cm~s$^{-1}$ is the wind velocity, $\Omega$ is the 
angular velocity of the binary, and $\zeta\sim 1$ is a numerical factor. 
Three X-ray binaries have been classified as massive ones with black hole 
companions: Cyg~X-1, LMC~X-1, and LMC~X-3 (see Tanaka \& Lewin 1995). 
They have orbital periods $P=2\pi/\Omega=5.6$, 4.2, and 1.7~d, respectively.
Hence, 
$\bar{l}\approx 1.5 \zeta P^{-1}(M/{\rm M}_\odot)(w/10^8)^{-4}r_gc \sim l_*$ 
in these objects, that implies marginal formation of a small-scale disc.
Whether the disc forms or not depends on the precise value of 
angular momentum which is difficult to calculate because the details 
of wind accretion are not fully understood (note the strong dependence
of $\bar{l}$ on the wind velocity, $w$). The fact that the objects are luminous
indicates that a disc does form. 

Small-scale accretion discs may also form around massive black holes in 
active galactic nuclei (AGN). Among possible sources of the accreting gas 
in AGN are star-star collisions and tidal disruption of stars by the BH 
(Hills 1975). The accretion flow is then likely to have a modest angular
momentum.

The regime $l\simgt\lmin$ can be widespread among luminous black-hole
sources. Note that only a small number of wind-fed BH binaries and galactic 
nuclei are active, and most of the unseen objects may accrete in the 
spherical regime $l<\lmin$ with a low luminosity. The observed bright 
sources with $l>\lmin$ are then in the `tail' of the distribution of objects 
over angular momentum, $\dd N/\dd l$. If this distribution falls steeply 
towards high $l$, most of the bright sources should be
near the threshold for disc formation. 

In this paper, we construct an axisymmetric model of accretion with 
$l\simgt\lmin$. The outline of the model is as follows.
The accretion flow starts at large distances where it is almost spherical;
the gas is Compton cooled and it falls freely towards the BH. 
Near the BH, the rotation deflects the infall from the radial direction and
a disc-like caustic appears in the defocused flow. Here the flow liberates 
energy in inelastic collision and then proceeds via the thin disc into the 
BH. Such a disc is drastically different from its standard counterpart
as regards the dynamics, the mechanism of energy release and the emission 
spectrum.

The paper is organised as follows. In Section~2, we study the axisymmetric
free fall in the gravitational field of a Schwarzschild BH and discuss the 
caustic in the equatorial plane (the collision ring) where the accretion 
disc is formed. In Section~3, we write down the basic equations of the disc.
In Section~4, we solve the equations numerically, obtain a model for the disc 
structure, and find the efficiency of accretion. Most of the energy is released 
quite close to the event horizon, at $\sim 2 r_g$, and the problem requires 
exact relativistic treatment. All the calculations are relativistic (assuming 
a non-rotating BH). For comparison, in Section 5, we also formulate and compute
the analogous Newtonian problem. The results and issues for further study are
discussed in Section~6.

%##########################################################################

\section{Formation of a small-scale disc}

Let us consider an axisymmetric accretion flow slowly rotating 
around the polar axis, so that the non-radial velocity $v_\varphi\ll v_r$ 
at $R\gg r_g$ (an asymptotically spherical inflow). As a result 
of Compton cooling by the inner source, the inflow becomes super-sonic inside 
the Compton radius, $\RC\sim 10^4 r_g$ (see e.g. Illarionov \& Kompaneets 
1990). At $R<\RC$, the matter accretes along ballistic parabolic trajectories. 
The freely falling flow in Newtonian gravity 
is discussed in Illarionov \& Beloborodov (2000, hereafter IB). 
Here we need the relativistic equations since the typical radius at which 
accretion switches from the spherical to a disc-like regime, 
$R\sim l^2/r_gc^2$, is comparable to $r_g$ in our problem. Hereafter label 
`$\infty$' corresponds to radii $R\gg r_g$ (but $R<\RC$).

\subsection{Free fall}

Consider a free-fall trajectory in Schwarzschild 
$(t,R,\theta,\varphi)$-coordinates. Let $\lambda$ be proper time along the 
trajectory. Denote the orbital angular momentum by ${\bf l}$; its projection 
onto the polar axis is
\beq
\nonumber
l_z=l\sin\theta_\infty.
\eeq
A parabolic trajectory (with specific orbital energy $E=c^2$) 
is described by equations (see e.g. Misner, Thorne \& Weeler 1973)
\be
   R^2\frac{\dd R}{\dd\lambda}=-\sqrt{r_gc^2R^3-l^2R^2(1-\frac{r_g}{R})},
\ee
\be 
   R^2\frac{\dd\theta}{\dd\lambda}=l_z\sqrt{\ctg^2\theta_\infty-\ctg^2\theta},
\ee
\be
   R^2\frac{\dd\varphi}{\dd\lambda}=\frac{l_z}{\sin^2\theta}.
\ee
The only difference from the corresponding Newtonian equations is 
the presence of the additional factor $(1-r_g/R)$ in equation (1).
One can solve equations (1-3) for the trajectory $\theta(R),\varphi(R)$,
\be
 \cos\theta=\cos\theta_\infty\cos\psi, 
\ee
\be
 \sin(\varphi-\vpi)=\frac{\sin\psi}{\sin\theta},
\ee
\be
 \psi(R)=\int_R^\infty\frac{l\,\dd R}{\sqrt{r_gc^2R^3-l^2R^2(1-r_g/R)}}. 
\ee 
The $\psi$ has the meaning of an orbital angle.
In the Newtonian limit, $r_g/R\rightarrow 0$, equation (6) gives 
$\cos\psi=1-l^2/GMR$ (eq.~11 in IB). 

The accreting matter streams from a sphere $S_\infty$ along the ballistic 
trajectories. The flow is assumed to be symmetric about the equatorial plane.
In this plane, a streamline has to collide with its symmetric 
counterpart (see Fig.~1). The loci of collisions form a 
two-dimensional caustic, the collision ring.

We assume that there are no intersections of the sreamlines before they
reach the equatorial plane. It is so if the Jacobian of the transformation 
$\theta_\infty,\varphi_\infty\rightarrow \theta(R),\varphi(R)$
stays positive along the streamlines.
Taking into account that $\partial\theta_R/\partial\varphi_\infty=0$ and
$\partial\varphi_R/\partial\varphi_\infty=1$ for any axisymmetric flow,
the condition of positive Jacobian reads
\beq
\nonumber
  \frac{\dd\cos\theta_R}{\dd\cos\theta_\infty}=\cos\psi
         -\cos\theta_\infty\sin\psi\frac{\dd\psi}{\dd\cos\theta_\infty}>0.
\eeq
It is satisfied if $\cos\thi\dd\psi/\dd\cos\theta_\infty<0$.
Since $\psi$ increases with increasing $l$, the condition can be rewritten as 
\be
 \frac{\dd l}{\dd\sin\theta_\infty}>0. 
\ee 
Hereafter we assume that condition (7) is satisfied.

%%%%%%%%%%%%%%%%%%%%%%%%%%%%%%%%%%%%%%%%%
\begin{figure}
\begin{center}
\leavevmode
\epsfxsize=8.4cm
\epsfbox{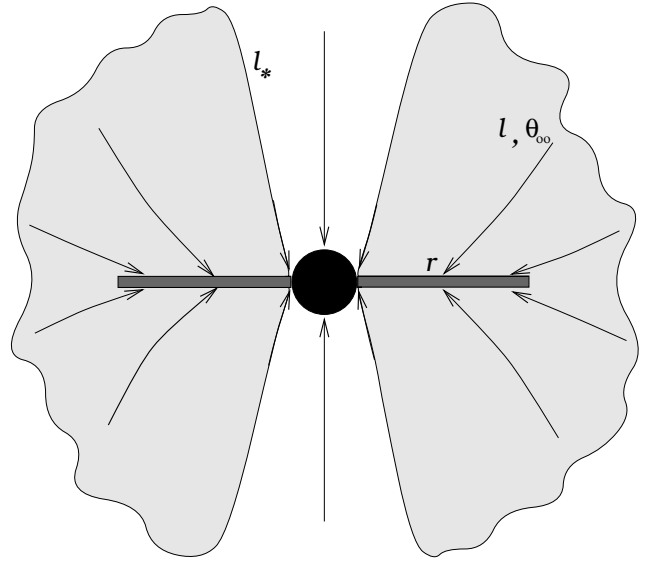}
\end{center}
\caption{ Schematic picture of the disc formation.
The inflow through $S_\infty$ has angular momentum $l$ 
increasing towards the equatorial plane, $\dd l/\dd\sin\thi>0$.
The collision radius, which is a growing function of $l$, monotonically 
increases as $\theta_\infty$ approaches $\pi/2$. 
The shadowed parts of the accretion flow collide
outside $r_g$ and form a couple of radiative shocks which sandwich the
collision ring. The other part of the flow (at small polar angles where
$l<l_*$)
is directly swallowed by the black hole with a low radiative efficiency.
}
\end{figure}
%%%%%%%%%%%%%%%%%%%%%%%%%%%%%%%%%%%%%%%%%

\subsection{The collision ring}

Streamlines coming from $S_\infty$ with $\thi\neq\pi/2$ collide in the 
symmetry plane and define a mapping $\vpi,\thi\rightarrow r,\varphi$. 
Here $(r,\varphi)$ are polar coordinates in the equatorial plane. 
(The $r$ has the same meaning as $R$; we use different notation 
to emphasize that $r$ measures radius in the plane.)
The mapping is given by equations (4-6) with $\theta=\pi/2$.
Equations (4) and (5) say that a streamline reaches the equatorial plane at
$\psi=\pi/2$ and executes 1/4 of the full turn around the polar axis, 
$\varphi=\vpi+\pi/2$. A circle $\thi=const$ at 
$S_\infty$ transforms into a circle $r=const$ with the $\pi/2$ rotation.
The radius of collision is found from $\psi=\pi/2$,
\be
   \int_r^\infty\frac{l\dd R}{\sqrt{r_gc^2R^3-l^2R^2(1-r_g/R)}}=\frac{\pi}{2}.
\ee
Here $l(\thi)=l(\pi-\thi)$ is the angular momentum of the colliding 
streamlines. Equation (8) involves an elliptic integral. 
At $r\gg r_g$, it is simplified and yields the Newtonian relation, $r=l^2/GM$.
The relativistic effects do not crucially alter the Newtonian expression
even at $r\sim r_g$. 
The solution to equation (8) can be written as an expansion which, in the 
leading order, yields
\begin{eqnarray}
\nonumber
   r=\frac{l^2}{GM}-\frac{(10-3\pi)}{4}r_g\approx \frac{l^2}{GM}-\frac{r_g}{7}.
\end{eqnarray}
This equation reproduces the exact solution $r(l)$ with high accuracy
(the maximum error $\simlt 0.5$ per cent at $r=r_g$).

The collision occurs outside the event horizon if $l$ 
exceeds a minimum value, $\lmin$, such that $r(\lmin)=r_g$.
From equation (8) one finds $\lmin\approx 0.754r_gc$. 
(In the Newtonian approximation, $\lmin=r_gc/\sqrt{2}\approx 0.707 r_gc$.) 
A streamline with $l<\lmin$ goes directly into the BH (see Fig.~1).
A streamline with $0.754<l/r_gc<2$ would also end up in the BH 
(see e.g. Novikov \& Frolov 1989). The collision in the equatorial plane, 
however, happens before that, at $r\simlt 8r_g$.

Streamlines starting at $\theta_\infty\rightarrow\pi/2$ have 
the largest angular momentum, $l\rightarrow l_0$, and 
define the outer edge of the collision ring, $r_0\approx l_0^2/GM-r_g/7$. 
 
A collisionless shock forms on each side of the ring. The accretion picture
shown in Fig.~1 assumes that the shock is radiative and pinned to the ring;
we study this condition in Section~2.3.
The free-fall equations (1-3) give the three components of the flow
velocity in the preshock region;
note that the azimuthal velocity is about the radial one.

\subsection{The pinned shocks and vertical mixing}

The vertical component of the plasma velocity is reduced at the shock front by
a factor of $\xi\geq 4$, while the horizontal velocity remains unchanged
(the horizontal motion towards the BH can stay super-sonic). 
The dissipated kinetic energy goes mostly to 
ion heating and some fraction $\epsilon$ may go to electron heating 
(studies of interplanetary and 
interstellar shocks indicate $\epsilon\sim 0.2-0.3$, see Draine \& McKee 1993). 
The electrons immediately (at the shock front) radiate the received energy
via inverse Compton scattering, producing hard gamma-rays (Shapiro \& Salpeter 
1975). The gamma-rays are converted into $e^\pm$ pairs and increase the 
downstream electron density by a factor of $z\simgt 1$. The cooled downstream 
$e^\pm$ then take the energy from the ions. 

The radiative capability of the shocked plasma depends on 
$\epsilon$ and also on the rate of energy exchange between the downstream
electrons and protons. The minimal e-p coupling is provided by Coulomb 
collisions: the electrons of temperature $T_e$ cool the protons
of temperature $T_p$ on a time-scale (Landau \& Lifshitz 1981),
\be
  t_{\rm Coul}=\frac{\sqrt{\pi/2}m_p\theta_e^{3/2}}{m_ec\sT zn\ln\Lambda}
             \approx 17 \frac{T_e^{3/2}}{zn} {\rm s},
\ee
where $n$ is the proton density, and a Coulomb logarithm 
$\ln\Lambda=15$ and 
$\theta_e\equiv kT_e/m_ec^2<1$ are assumed. The e-p coupling can be enhanced 
by collective plasma effects, so that the proton cooling time-scale is 
$\tep=\chi^{-1} t_{\rm Coul}$ where $\chi>1$.

The shock is radiative and held-down to the equatorial plane if $\tep$ is 
shorter than $\xi r/\vff$ where $\vff=c(r_g/r)^{1/2}$ is the typical 
free-fall velocity. The downstream density can be estimated in terms of the 
accretion rate $\dMt$: $n\approx \xi\dMt/4\pi r^2m_p\vff$.
The condition that the shock is pinned to the equatorial plane then reads
\be
   \frac{\vff\tep}{\xi r}=\frac{\sqrt{2\pi}m_p\theta_e^{3/2}}
                              {z\ln\Lambda\, m_e\xi^2\chi\dm}<1,
\ee
where $\dm=\dMt c^2/\LE$ is the dimensionless total accretion rate and
$\LE=2\pi r_g m_p c^3/\sT$ is the Eddington luminosity. 

The downstream electron temperature can be found from the energy balance,
\be
  \frac{zT_e}{\tC}=\frac{T_p}{\tep},
\ee
where $\tC=3m_ec/8\sT U$ is the Compton cooling time-scale and $U$ is the 
radiation energy density. For a radiative shock, the proton thermal energy 
is transformed into a radiation flux 
$F\sim(3/2)nkT_p\vff/\xi$. It yields an estimate,
$U\sim 2F/c\sim 3nkT_p\vff/c\xi$, and hence
\beq
\nonumber
  \tC\approx \frac{m_ec^2\xi}{8\sT nkT_p\vff}.
\eeq
The energy balance (11) then yields 
\be
  \theta_e\approx\left(\frac{\ln\Lambda\, m_e}{\sqrt{2\pi}m_p}
                      \frac{\xi\chi c}{4\vff}\right)^{2/5}
          \approx 0.1\left(\frac{\xi\chi c}{4\vff}\right)^{2/5}. 
\ee
$\vff\propto r^{-1/2}$ and hence $\theta_e\propto r^{1/5}$. 
The temperature weakly depends on parameters.

The hot downstream layer radiating most of the proton heat 
has optical depth $\tT\approx (\vff/\xi)\tep zn\sT$. It satisfies the relation,
$\tT\theta_e\approx 0.1$ (as follows from eqs.~9 and 12), i.e, $\tT\sim 1$
for typical $\theta_e\sim 0.1$. With more exact approach, the unsaturated 
Comptonization in the hot layer involves radiative transfer with multiple 
scatterings, and $\tT\sim 1$ corresponds to a Kompaneets' parameter $y\sim 1$
(see e.g. Rybicki \& Lightman 1979). 
In deeper layers, at $\tT>1$, the downstream temperature 
(both $T_p$ and $T_e$) falls off sharply (see Zel'dovich \& Shakura 1969,
Shapiro \& Salpeter 1975 for a similar shock on the surface of a neutron star).

Using equation (12), we rewrite the condition (10) as
\be
   \dm>\dm_0\equiv \frac{0.5}{z\chi^{2/5}}\left(\frac{4}{\xi}\right)^{7/5} 
                         \left(\frac{c}{\vff}\right)^{3/5}
                  \approx \frac{0.05}{z\theta_e}\left(\frac{4}{\xi}\right)
                                                \left(\frac{c}{\vff}\right).
\ee
We are interested here in sub-Eddington sources (with luminosities 
$L<\LE$) so that the free-fall approximation can be used for the super-sonic
infall. It requires $\dm<\dmE=\eta^{-1}$ where $\eta$ is the radiative 
efficiency of the disc (see Section~4 and Fig.~6). 
In the range $\dm_0<\dm < \eta^{-1}$ we have sub-Eddington 
accretion with radiatively efficient shock pinned to the equatorial plane. 

This pattern of accretion is different from the previously considered situation 
$\dm\simlt\dm_0$ (see e.g. Igumenshchev, Illarionov \& Abramowicz 1999).
In that case, the plasma is unable to cool fast and supports quasi-spherical 
shocks away from the collision ring. The hydrodynamic simulations of Igumenshchev 
et al. (1999) show that the shock tends to shrink towards the ring of radius 
$l_0^2/GM$ when $\dm$ increases. 

We hereafter assume $\dm>\dm_0$ and have a free-falling flow until it reaches 
the thin disc of shocked plasma. The disc has a finite half-thickness $H$ 
supported by the pressure $p_d\sim \rho_dc_s^2$ against the ram pressure of 
the infall, $p_{\rm ff}\sim\rho_{\rm ff}\vff^2$. Here $\rho_d$ and $c_s$ are 
the density and the sound speed in the midplane of the disc, and 
$\rho_{\rm ff}$ is the density of the infall. The pressure balance yields 
$\rho_d/\rho_{\rm ff}\sim (\vff/c_s)^2$. The regime of pinned shock 
implies efficient radiative cooling and we have $c_s\ll\vff$ and 
$\rho_d\gg\rho_{\rm ff}$. The accretion velocity in the disc is comparable to 
$\vff$ (see Sections~3 and 4) and from matter conservation we estimate 
$H/r\sim \rho_{\rm ff}/\rho_d\sim (c_s/\vff)^2$. 

At a given radius, the disc is composed of matter that entered
the disc at larger radii with different horizontal velocities.
A strong turbulence is likely to develop in the disc under such conditions. 
The turbulence can mix the disc in the vertical direction on a time-scale
$\sim H/c_s$ which is $\sim c_s/\vff$ shorter than the accretion time-scale,
$t_a\sim r/\vff$. In the mixing  process, the random motions above the level 
$\sim \rho_d c_s^2$ are converted into heat and the disc acquires 
a vertically-averaged horizontal velocity (with both radial and azimuthal 
components). The released heat is radiated away, contributing to the total 
luminosity.

When absorbing the infalling matter,
the disc also absorbs horizontal momentum and its velocity changes. 
The process of disc-infall inelastic collision 
is governed by the corresponding law of momentum conservation (see Section~3). 
The disc is a `sticky' caustic in the accretion flow and it resembles
the accretion line in the Bondi-Hoyle-Lyttleton (BHL) problem 
(see Bondi \& Hoyle 1944). We discuss this analogy in more detail below.

%##########################################################################

\section{Fast inviscid accretion disc}

The shocked turbulent gas continues to accrete into the BH through the thin 
disc. We will show that if the inflow angular momentum is smaller than some 
specific value then accretion in the disc occurs fast: the gas spirals into 
the BH on the free-fall time-scale. In contrast to the standard viscous disc,
the radial velocity is much larger than the sound speed. Therefore, 
turbulent viscosity has a negligible effect on accretion, i.e. the disc is 
inviscid in the horizontal directions (yet the turbulence efficiently mixes 
the disc in the vertical direction, see Section 2.3). Thereafter we solve 
the problem of steady disc accretion in the inviscid regime.
In illustrative calculations we assume solid body rotation of the inflow at 
$S_\infty$,
\be
  l(\thi)=l_0\sin\thi, \qquad \bar{l}_z=\int_0^{\pi/2}l\;\sin^2\thi\dd\thi
         =\frac{2}{3}l_0.
\ee 
Here $\bar{l}_z$ is the average specific angular momentum of the inflow.

\subsection{Accretion rate}

The asymptotically spherical inflow at $S_\infty$ has a total 
accretion rate $\dMt$ and a homogeneous distribution 
$\dd\dM/\dd\Omega_\infty=\dMt/4\pi$ where 
$\dd\Omega_\infty=\dd\vpi\dd\cos\thi$. 
All the matter impinging the disc finally accretes into the BH. Hence,
the disc accretion rate at a radius $r$, $\dM(r)$, equals the rate 
of matter supply to the disc at all $r^\prime>r$, which in turn equals 
the rate of accretion with $l^\prime>l$ through the equatorial segment 
$\sin\thi^\prime>\sin\thi$ at $S_\infty$ (see Fig.~1),
\be
 \dM(r)=\dM_{>l}=\dMt\cos\theta_\infty(l).
\ee

In the case of solid body rotation at $S_\infty$ (eq.~14), 
we have the relation $\sin\theta_\infty=l/l_0$ and hence
\be
\dM(r)=\dMt\sqrt{1-\left(\frac{l}{l_0}\right)^2}.
\ee
Here $l$ is a function of $r$, see equation (8).
With the approximate relation $l^2\approx GMr$, we get
\be
  \dM(r)\approx \dMt\sqrt{1-\frac{r}{r_0}}.
\ee

\subsection{Angular momentum}

The specific angular momentum of a steady inviscid disc, 
$j(r)$, equals the average angular momentum,
$\bar{l_z^\prime}$, of matter absorbed by the disc at $r^\prime>r$. 
(It follows from the conservation of the accreting mass and conservation of 
the total accreted angular momentum. The radiation carries away a small 
fraction of the disc momentum, see Appendix A; hereafter we neglect these 
losses.)  We get,
\be
 j(r)=\frac{-1}{\dM(r)}\int_r^{r_0} 
   l_z^\prime(r^\prime)\frac{\dd\dM}{\dd r^\prime} \dd r^\prime=
 \frac{-1}{\dM_{>l}}\int_{l}^{l_0}
 l_z^\prime\frac{\dd\dM}{\dd l^\prime}\dd l^\prime.
\ee
Here $l_z^\prime=l^\prime\sin\thi^\prime$ and $\dM_{>l}$ is given by 
equation (15).
The infall has $l_z(r)<j(r)$ and decelerates the disc rotation: 
$j(r)$ decreases inwards.

In the case of solid body rotation at $S_\infty$, we have $\sin\thi=l/l_0$, 
$l_z=l^2/l_0$ and $\dd\dM/\dd l=-\dMt(l/l_0)(l_0^2-l^2)^{-1/2}$.
Working out the integral (18) then yields
\be
 j(r)=\frac{2}{3}l_0+\frac{1}{3}l_z(r)=\frac{2}{3}l_0+\frac{l^2(r)}{3l_0},
\ee
(see Fig.~2). With the approximate relation $l^2\approx GMr$, we get
\be
  j(r)\approx l_0\left(\frac{2}{3}+\frac{r}{3r_0}\right).
\ee

\subsection{Radial equation}

The matter in the disc is not in free fall because it absorbs the infalling 
matter. The infall accelerates the accretion in the disc (the law of momentum 
conservation is given in Appendix~A).
The equation for the disc radial velocity $u^r$ reads,
\be
 \frac{\dd u^r}{\dd r}=\frac{\dd\dM}{\dd r}\frac{(\hat{u}^r-u^r)}{\dM(r)}
-\frac{r_gc^2}{2r^2u^r}\left(1-\frac{j^2}{\jK^2}\right).
\ee
Here $j$ is given by equation (18) and
\be
  \jK^2\equiv \frac{r^2r_gc^2}{2r-3r_g}. 
\ee
At $r>(3/2)r_g$, 
$\jK^2$ has the meaning of the squared angular momentum for circular Keplerian 
rotation. At $r<(3/2)r_g$, such motion is not possible in Schwarzschild
geometry, which is reflected by the fact that $\jK^2<0$. 
The $\hat{u}^r$ in equation (21) is the radial velocity, $\dd r/\dd\lambda$,
of matter impinging the disc at radius $r$ (see eq.~1),
\be
 \hat{u}^r=-\sqrt{\frac{r_gc^2}{r}-\frac{l^2}{r^2}\left(1-\frac{r_g}{r}\right)},
\ee
with $l$ being related to $r$ via equation (8).

Note that equation (21) is formally identical to the corresponding
equation in Newtonian gravity (considering $u^r$ as a normal Newtonian radial 
velocity). The relativistic effects come into the problem through $\jK^2$,
$\hat{u}^r$, and the relation (8) between $l$ and $r$.

The first term on the right-hand-side of the radial equation is a result of
the disc-infall interaction. 
Assume for a moment that inside a radius $r_1$ there is no matter supply to 
the disc. Then at $r<r_1$ the disc matter is freely falling with constant
$\dM$, $j$, and $E$, where $E$ is the orbital energy given by equation (24)
below. In the presence of matter supply, 
$\dd\dM/\dd r\neq 0$, both $j$ and $E$ change with radius.

The radial equation (21) is similar to the BHL equation describing gas motion 
along the accretion line (see eq.~8 in Bondi \& Hoyle 1944). 
The only difference is that we have an additional term proportional to 
$j^2/\jK^2$. This term reflects the fact that the disc matter is rotating and 
a repulsive centrifugal force appears. The centrifugal force dominates over 
gravity if $j>\jK$ and it can stop accretion. Therefore, a steady inviscid 
disc may form only in accretion flows with limited net angular momentum 
(see Section~3.5).

The differential equation (21), like the BHL equation, does not admit 
analytical solution. We solve numerically equation (21) combined with 
the formulae (15), (18), (22), (23) for $\dM$, $j$, $\jK^2$, $\hat{u}^r$, 
respectively, and account for the relation (8) between $l$ and $r$.
The outer boundary condition is $u^r(r_0)=\hat{u}^r$. 
The solution $u^r(r)$, $j(r)$, $\dM(r)$
yield a closed description of the fast inviscid disc. The model is fully 
determined by the inflow angular momentum $l(\theta_\infty)$.

\subsection{Energy release}

Once the disc dynamics is solved and $u^r(r)$, $j(r)$, $\dM(r)$ are known, 
one can compute the energy liberated by the disc-infall collision.
The orbital energy of the disc matter $E=-u_t$ (see Appendix A) equals $c^2$
at the outer edge of the disc and decreases inwards. From $u_iu^i=-c^2$ we have
\beq
 E^2=c^2\left(u^r\right)^2+V^2, \quad
 V^2(r)\equiv c^2\left(c^2+\frac{j^2}{r^2}\right)
                 \left(1-\frac{r_g}{r}\right).
\eeq
The material absorbed by the fast disc at $r^\prime>r$ and accreted down to 
$r$ liberates a luminosity $L_{>r}=\dM(r)[c^2-E(r)]$.
The total energy released by the disc is 
\be
 L=\dM_d\left[c^2-E(r_g)\right].
\ee
Here $\dM_d=\dM(r_g)$ is the total accretion rate through the disc and 
$c^2-E(r_g)$ is the binding energy of gas swallowed by the BH.
The energy liberated per unit radius is
\be
 \frac{\dd L}{\dd r}=\left|\frac{\dd\;\,}{\dd r}
  \left[\dM(r)\left(c^2-E(r)\right)\right]\right|.
\ee

\subsection{The limitation on the inflow angular momentum}

The essential condition for an inviscid regime of accretion is that 
there is no turning point in the disc ($u^r<0$ everywhere). 
It requires $E>V$ (see eq.~24) where $V(r)$ is the effective potential
for radial motion (Landau \& Lifshitz 1975; see Misner et al. 1973, ch.~25 
for detailed discussion). The potential has a pit at small $r$ and
matter may fall into the BH. With increasing angular
momentum, there appears a centrifugal barrier enclosing the pit and 
preventing accretion.

The main quantity determining the disc dynamics is the angular momentum
$j(r)$. The no-turning-point condition $E>V$ can be expressed as
\beq
  j(r)<j_t(r)\equiv\frac{cr}{\sqrt{r-r_g}}\sqrt{r_g
                          +\left(\frac{E^2}{c^4}-1\right)r},
\eeq
where $j=j_t$ corresponds to $E=V$.
The disc orbital energy is not much below the parabolic value $c^2$ and,
in the first approximation, $j_t$ can be estimated assuming $E=c^2$,
\beq
j_{t0}=cr\sqrt{\frac{r_g}{r-r_g}},
\eeq
which is slightly larger than $j_t$. The $j_{t0}$ reaches a minimum of
$2r_g c$ at $r=2r_g$ (see Fig.~2). The fast disc can exist only with
$j(2r_g)<2r_gc$. Note that $j(r)>\bar{l}_z$ at any $r$, where $\bar{l}_z$ 
is the average angular momentum of the inflow.
Therefore, the necessary condition for the fast-disc regime reads
\beq
\bar{l}_z<2r_gc.
\eeq
This limitation does not exclude distributions $l(\thi)$ with a tail of large 
$l$ at $\thi\rightarrow\pi/2$ (and the inviscid disc can have an infinitely 
large radius). However, most of the accreting mass is required to have a 
modest angular momentum.

If the condition (29) is not satisfied, there is no stationary solution to 
the inviscid problem. Instead a viscous disc forms which spreads out
to large radii (Kolykhalov \& Sunyaev 1979). The transition to the viscous 
regime is rather complicated, showing unstable time-dependent behaviour, 
and it will be studied in a separate paper (Beloborodov \& Illarionov, 
in preparation). Here we restrict our consideration to the fast inviscid 
regime.

%%%%%%%%%%%%%%%%%%%%%%%%%%%%%%%%%%%%%%%%%
\begin{figure}
\begin{center}
\leavevmode
\epsfxsize=8.4cm
\epsfbox{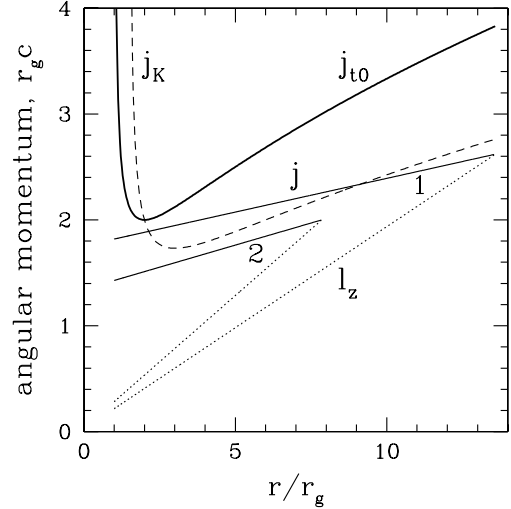}
\end{center}
\caption{ 
Two characteristic angular momenta of the problem, $j_{t0}(r)$ (eq.~28) and 
$\jK(r)$ (eq.~22), are shown by the heavy and dashed curves, respectively.
The necessary condition for the fast disc regime is that the disc angular 
momentum $j(r)<j_{t0}(r)$ everywhere. The bottleneck for accretion is at $2r_g$.
The solid lines show $j(r)$ in the model computed in Section~4 for two cases:
$l_0=\lcr\approx 2.62r_gc$ (marked 1)  and $l_0=2r_gc$ (marked 2).
The angular momentum of matter impinging the disc, $l_z(r)$,
is shown by dotted lines. 
}
\end{figure}
%%%%%%%%%%%%%%%%%%%%%%%%%%%%%%%%%%%%%%%%%

%#############################################################################

\section{Numerical model}

We now compute the illustrative model of a fast disc formed by the inflow
with angular momentum (14); then $l_0=(3/2)\bar{l}_z$ is the only parameter 
of the problem. The limitation (29) requires $l_0<3r_gc$. The exact critical
value of $l_0$ is found below.

\subsection{Dynamics}

The disc accretion rate $\dM(r)$ and its angular momentum $j(r)$ are given 
by equations (16) and (19). 

The solution for the accretion velocity is shown in Fig.~3.
At $l_0=\lcr\approx 2.62r_gc$, the velocity profile touches zero: 
accretion is stopped by the centrifugal barrier at $\rcr\approx 2.14r_g$
($j$ touches $j_t$, see eq.~27, and it always stays smaller than $j_{t0}$,
see Fig.~2). 
The critical radius is near $2r_g$, close to the minimum of $j_{t0}$.
At $l>\lcr$ the fast inviscid disc cannot exist and hence its maximum radius 
is $r_{\rm max}=r_0(\lcr)\approx 13.6r_g \approx 27 GM/c^2$.

%%%%%%%%%%%%%%%%%%%%%%%%%%%%%%%%%%%%%%%%%
\begin{figure}
\begin{center}
\leavevmode
\epsfxsize=8.4cm
\epsfbox{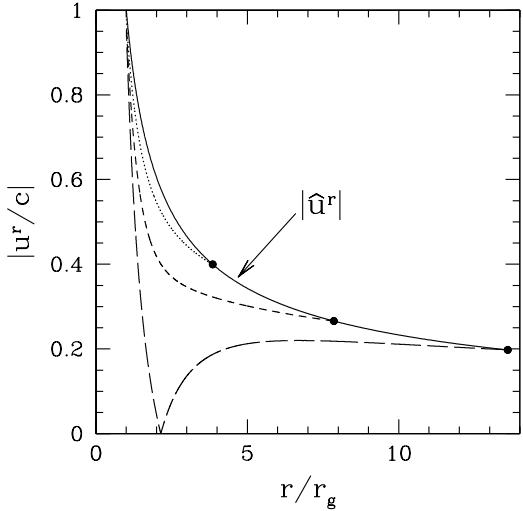}
\end{center}
\caption{ The radial velocity of the disc, $u^r$ (in units of $c$). The 
long-dashed, dashed, and dotted curves correspond to the cases 
$l_0=\lcr\approx 2.62r_gc$, $l_0=2r_gc$, and $l_0=\sqrt{2}r_gc$, respectively
(see eq.~14 for the definition of $l_0$). The outer radius of the disc,
$r_0\approx 2l_0^2/r_gc^2-r_g/7$, is shown by solid dots.
The radial velocity of matter impinging the disc, $\hat{u}^r$,
is shown by the solid curve (eq.~23).
}
\end{figure}
%%%%%%%%%%%%%%%%%%%%%%%%%%%%%%%%%%%%%%%%%

In Fig.~4, we show the trajectory of the spiraling gas in the disc.
In the critical case, the gas makes infinite number of revolutions 
at $\rcr$ before it falls into the BH. The disc surface density 
$\Sigma(r)=\dM(r)/2\pi r u^r\rightarrow \infty$ at $r\rightarrow\rcr$.

%%%%%%%%%%%%%%%%%%%%%%%%%%%%%%%%%%%%%%%%%
\begin{figure}
\begin{center}
\leavevmode
\epsfxsize=8.4cm
\epsfbox{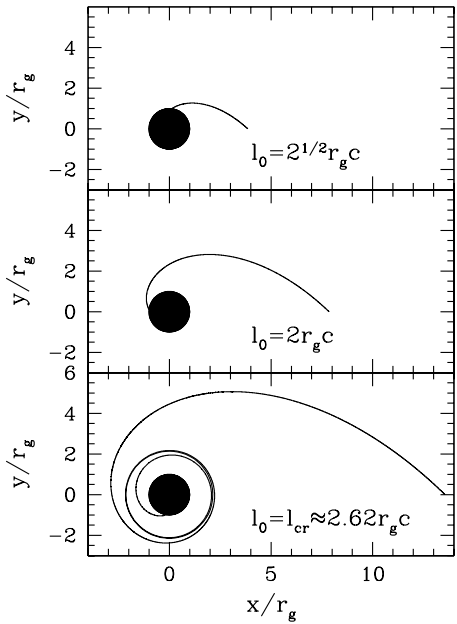}
\end{center}
\caption{The trajectory of gas motion in the disc, from the outer edge 
into the black hole. In the critical case (bottom panel) the trajectory makes an
infinite number of revolutions at the critical radius $\rcr\approx 2.14r_g$ before
falling into the black hole. 
}
\end{figure}
%%%%%%%%%%%%%%%%%%%%%%%%%%%%%%%%%%%%%%%%%

\subsection{The radiative efficiency}

In Fig.~5 we show the distribution of the liberated energy over radius,
$r\dd L/\dd r$. The energy release peaks at $\approx 2r_g$. 
The sharp peak in the critical case is caused by the minimum of $u^r$ 
(see Fig.~3). The relative velocity between the disc and the infall 
increases dramatically at $r=\rcr$ and the inelastic collision liberates 
more energy. Note that in the standard disc model 
(Novikov \& Thorne 1973) $r\dd L/\dd r$ peaks at $\approx 9r_g$.

The small size of the main emitting region implies that 
the released energy is partly captured into the BH. Accounting for
the capture effect, we evaluate the observed luminosity in Appendix~B,
\beq
 \frac{\dd L_{\rm obs}}{\dd r}=\frac{\dd L}{\dd r}\left[1-\kappa(r)\right].
\eeq
The capture is enhanced by the Doppler beaming of the disc radiation 
into the BH: the emitting material accretes fast. 
The effect is especially strong when $l_0$ is small (see Fig.~5). 
In the critical case, $l_0=\lcr$, the accretion is decelerated 
at $r\approx 2r_g$ and the luminosity capture is less efficient.

There are a few ways to define the radiative efficiency of accretion: 
(i) $\eta=L/\dMt c^2$, (ii) $\eta=L/\dM_d c^2$, 
(iii) $\eta=L_{\rm obs}/\dMt c^2$, and (iv) $\eta=L_{\rm obs}/\dM_d c^2$. 
We choose the first definition as a basic one 
and compute the other three for comparison. The results are shown in Fig.~6.

%%%%%%%%%%%%%%%%%%%%%%%%%%%%%%%%%%%%%%%%%
\begin{figure}
\begin{center}
\leavevmode
\epsfxsize=8.4cm
\epsfbox{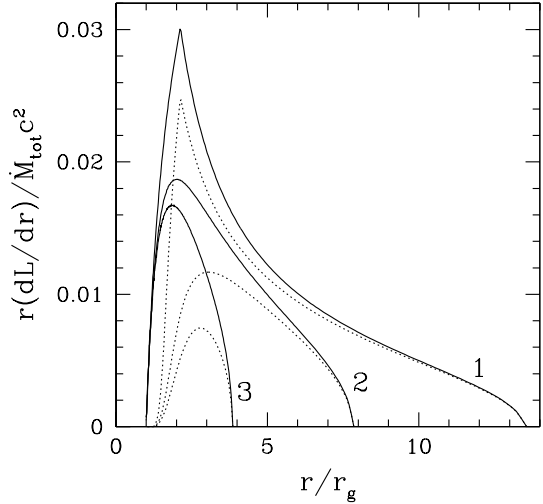}
\end{center}
\caption{ The radial distribution of the disc luminosity: solid curves -- 
the released luminosity (eq.~26), dotted curves -- the observed luminosity 
corrected for the gravitational capture by the black hole (eq.~30). 
Labels 1, 2, and 3 correspond to the cases $l_0=\lcr\approx 2.62 r_gc$,
$l_0=2r_gc$, and $l_0=\sqrt{2}r_gc$, respectively. 
}
\end{figure}
%%%%%%%%%%%%%%%%%%%%%%%%%%%%%%%%%%%%%%%%%

%%%%%%%%%%%%%%%%%%%%%%%%%%%%%%%%%%%%%%%%%
\begin{figure}
\begin{center}
\leavevmode
\epsfxsize=8.4cm
\epsfbox{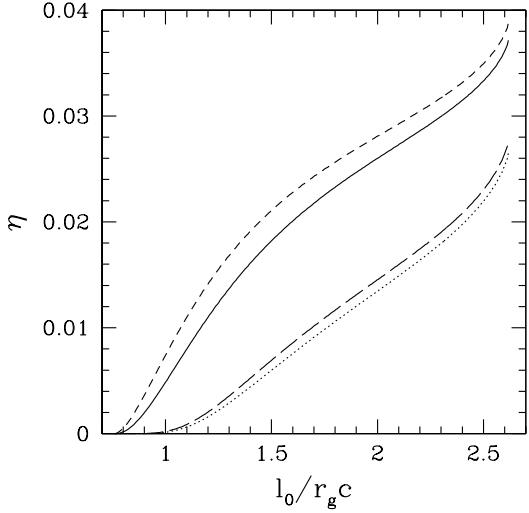}
\end{center}
\caption{ The efficiency of accretion, $\eta$, versus $l_0$. The four curves
correspond to different definitions of efficiency: solid curve --
$\eta=L/\dMt c^2$, dashed curve -- $\eta=L/\dM_d c^2=1-E(r_g)/c^2$, 
dotted curve -- $\eta=L_{\rm obs}/\dMt c^2$, long dashed curve -- 
$\eta=L_{\rm obs}/\dM_d c^2$. 
Here $L$ is the power released by the disc (eq.~25), $L_{\rm obs}$ is the 
luminosity corrected for the capture by the black hole,
$\dMt$ is the total accretion rate (including the polar inflow that plunges 
into the event horizon before crossing the equatorial plane, see Fig.~1),
and $\dM_d$ is the accretion rate through the disc.
}
\end{figure}
%%%%%%%%%%%%%%%%%%%%%%%%%%%%%%%%%%%%%%%%%

The maximum efficiency, $\eta\approx 0.0372$, is achieved in the critical case, 
$l_0=\lcr$. 
It is smaller than the efficiency of the standard disc around a Schwarzschild
black hole, $\eta=1-2\sqrt{2}/3\approx 0.0572$, however it is comparable to
this value. We conclude that the fast inviscid disc is quite an efficient 
regime of accretion.

%#############################################################################

\section{Newtonian fast disc}

In this section, we formulate and solve the fast disc problem in Newtonian 
gravity; the importance of the relativistic effects is well seen when 
comparing the BH fast disc with its Newtonian counterpart. 
The natural units of length and velocity in the relativistic problem are 
$r_g$ and $c$. The corresponding units in the Newtonian problem are the
accretor radius $R_*$ and the free-fall velocity
\beq
\nonumber
  u_*=\left(\frac{2GM}{R_*}\right)^{1/2}.
\eeq

\subsection{Dynamics and energy release}

The Newtonian disc fed by matter coming from $S_\infty$ with angular 
momentum $l(\thi)$ is described by the same dynamical equations (15), (18), 
and (21) as the relativistic disc. The only difference appears in
the expressions for $r(l)$, $\jK^2(r)$, and $\hat{u}^r$,
\be
  r=\frac{l^2}{GM}, \qquad \jK^2=GMr, \qquad \hat{u}^r=-\sqrt{\frac{GM}{r}}
\ee 
(compare with eqs.~8, 22, and 23). The major change is the equation for $\jK^2$, 
which makes the disc dynamics crucially different from the relativistic case.

The specific orbital energy in Newtonian theory is
\be
E(r)=\frac{(u^r)^2}{2}+V(r), \qquad V(r)\equiv -\frac{GM}{r}+\frac{j^2}{2r^2}.
\ee
$E=0$ for the parabolic infall above the equatorial plane; the 
binding energy of the disc matter equals $-E(r)$.

The energy liberated in the disc outside a given radius $r$ equals
$L_{>r}=-E(r)\dM(r)$. The total luminosity of the disc is given by 
\be
 L=-E(R_*)\dM_d,
\ee
where $\dM_d=\dM(R_*)$ is the accretion rate through the disc and $-E(R_*)$ 
is the binding energy of matter that reaches the accretor. The energy 
liberated per unit radius is
\be
 \frac{\dd L}{\dd r}=\left|\frac{\dd\;\,}{\dd r}
  \left[-E(r)\dM(r)\right]\right|.
\ee

The effective potential for radial motion in the disc, $V(r)$, is given by 
equation (32).
The important difference from the BH case is that the pit at small $r$ is now 
absent ($V\rightarrow \infty$ at $r\rightarrow 0$) and the potential 
produces a stronger repulsive centrifugal force.
It leads to tighter constraints on the angular momentum of the fast disc 
regime. The condition that the disc has no turning point ($E>V$) in terms
of angular momentum reads,
\beq
 j(r)<j_t(r)=r\sqrt{2\left(\frac{GM}{r}+E\right)}.
\eeq
At $E=0$ it yields $j_{t0}=\sqrt{2}\jK$. The minimal value $j_{t0}=R_*u_*$
at $r=R_*$ implies the necessary condition for the disc
formation,
\beq
  \bar{l}_z<R_*u_*.
\eeq
This limitation is twice as stringent as that in the relativistic case (compare 
eq.~36 with eq.~29).

\subsection{Numerical model}

We now compute an illustrative numerical model for 
the same inflow at $S_\infty$ as in the relativistic problem, 
$\dd\dM/\dd\Omega_\infty=\dMt/4\pi$ and $l(\theta_\infty)=l_0\sin\thi$.
We solve numerically equation (21) coupled with expressions (17), (20), and (31).
The outer radius of the disc is given by $r_0=l_0^2/GM$, and the outer boundary
condition is $u^r(r_0)=\hat{u}^r(r_0)$. At the inner boundary, $r=R_*$, the 
accreting matter is absorbed by the central object.

%%%%%%%%%%%%%%%%%%%%%%%%%%%%%%%%%%%%%%%%%
\begin{figure}
\begin{center}
\leavevmode
\epsfxsize=8.4cm
\epsfbox{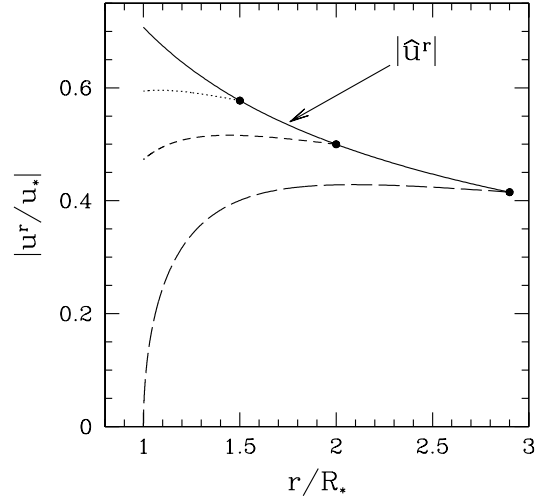}
\end{center}
\caption{ The radial velocity of the Newtonian fast disc, $u^r(r)=\dd r/\dd t$. 
The long-dashed, dashed, and dotted curves show the cases 
$l_0=\lcr\approx 1.204 R_*u_*$, $l_0=R_*u_*$, and $l_0=(\sqrt{3}/2)R_*u_*$,
respectively (see eq.~14 for the definition of $l_0$); 
$u_*\equiv (2GM/R_*)^{1/2}$. 
The outer radius of the disc, $r_0=2l_0^2/R_*u_*^2$, is shown by solid dots.
The radial velocity of matter impinging the disc, $\hat{u}^r$,
is shown by the solid curve (eq.~31).
}
\end{figure}
%%%%%%%%%%%%%%%%%%%%%%%%%%%%%%%%%%%%%%%%%

%%%%%%%%%%%%%%%%%%%%%%%%%%%%%%%%%%%%%%%%%
\begin{figure}
\begin{center}
\leavevmode
\epsfxsize=9.0cm
\epsfbox{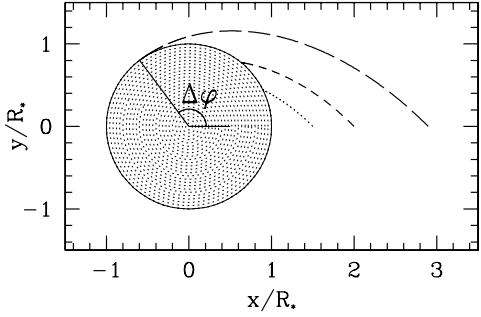}
\end{center}
\caption{The streamlines of the Newtonian fast disc, from the 
outer edge onto the surface of the accretor, $R_*$. 
The dotted, dashed, and solid curves show the cases $l_0=(\sqrt{3}/2)R_*u_*$, 
$l_0=R_*u_*$, and $l_0=\lcr\approx 1.204 R_*u_*$, respectively. 
The rotation angle, $\Delta\varphi=2.21$ is shown for the critical case.
}
\end{figure}
%%%%%%%%%%%%%%%%%%%%%%%%%%%%%%%%%%%%%%%%%

Example solutions are shown in Fig.~7-9.
Like the BH case, there exists a critical angular momentum, $\lcr$, such that
accretion flow with $l_0>\lcr$ is stopped by the centrifugal barrier. 
Numerically, we find $\lcr=1.204 R_*u_*$. 
The minimum angular momentum necessary for the disc formation is 
$\lmin=R_*u_*/\sqrt{2}$.  
The fast disc regime thus exists at $0.707<l_0/R_*u_*<1.204$.

At $l_0=\lcr$, the turning point first appears at the inner boundary of the 
disc, $r=R_*$. The disc streamlines then touch the accretor with 
$u^r\rightarrow 0$; the gas moving from the outer edge $r_0$ rotates by 
an angle $\Delta\varphi\approx 2.21$ (see Fig.~8). The critical behaviour of 
the BH disc is qualitatively different: $\rcr$ is
twice as large as the BH radius and gas rotates an infinite number of revolutions
(see Fig.~4). 

The maximum radius of the fast disc is achieved at $l_0=\lcr$, 
$\rmax/R_*=\lcr^2/GMR_*\approx 2.90$. Hence, the Newtonian fast disc is five 
times smaller than its relativistic counterpart 
($\rmax/r_g\approx 13.6$ in the BH case).
We conclude that the large extension of the BH fast disc was caused by 
purely relativistic effects, namely, by the pit in the effective potential.

We now compute the disc luminosity. First of all, note that the total energy 
released in the process of accretion onto a star of radius $R_*$ is 
$L_{\rm tot}=\dMt GM/R_*$. The bulk of the energy is released on the star 
surface, and only a fraction is liberated in the disc itself. 
In Fig.~9 we show the radial distribution of the disc luminosity.
In contrast to the BH case, the luminosity peaks at $r=R_*$ (compare with 
Fig.~5).

%%%%%%%%%%%%%%%%%%%%%%%%%%%%%%%%%%%%%%%%%
\begin{figure}
\begin{center}
\leavevmode
\epsfxsize=8.4cm
\epsfbox{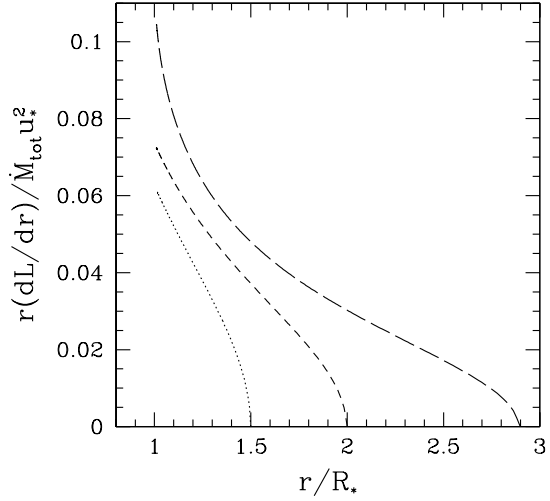}
\end{center}
\caption{ The radial distribution of the Newtonian fast disc luminosity in 
units of $\dMt u_*^2$.
The long-dashed, dashed, and dotted curves show the cases 
$l_0=\lcr\approx 1.204 R_*u_*$, $l_0=R_*u_*$, and $l_0=(\sqrt{3}/2)R_*u_*$, 
respectively. 
}
\end{figure}
%%%%%%%%%%%%%%%%%%%%%%%%%%%%%%%%%%%%%%%%%

The radiative efficiency of the disc is determined by the binding energy,
$E_b=-E(R_*)$ at the inner edge. The $E_b$ 
increases with $l_0$ and reaches its maximum at $l_0=\lcr$,
\beq
\nonumber
  E_b=\frac{GM}{R_*}\left[1-\frac{1}{2}\left(\frac{2\lcr}{3\lmin}
  +\frac{\lmin}{3\lcr}\right)^2\right]\approx 0.0572 u_*^2.
\eeq
We here accounted for the fact that $u^r=0$ at $r=R_*$ and used equation (19).
Amusingly, the critical binding energy of the Newtonian fast disc
in units of $u_*^2$ exactly coincides with the binding energy of the 
marginally stable Keplerian orbit around a Schwarzschild BH, 
$(c^2-E)/c^2=1-2\sqrt{2}/3\approx 0.0572$.

%##########################################################################

\section{Discussion}

\subsection{Special features of the fast disc regime}

There are two principle features of the accretion regime studied in this paper,
which make it different from previous models of rotating accretion flows
onto black holes:

(i) The Compton cooled quasi-spherical flow is falling freely until it 
collides with the small-scale disc. The disc is sticky: the shock
is radiative and pinned to the disc. 
This regime requires a sufficiently high accretion rate (see Section 2.3).
A similar behaviour (held-down shock at high $\dM$) was found for spherical 
accretion onto a neutron star (Shapiro \& Salpeter 1975). 

(ii) The small-scale disc has a modest angular momentum and a high 
radial velocity. The centrifugal barrier does not prevent accretion and
matter is swallowed by the black hole in a steady regime.
The effective throat of the black hole has a large radius of $2r_g$: 
the bottleneck for accretion is at $2r_g$ (see Fig.~2).
This relativistic effect makes the accretion 
easier and allows the fast disc radius to be as large as 
$\approx 13.6 r_g\approx 27GM/c^2$ (in Newtonian gravity it would be five
times smaller, see Section~5.2).

The fast disc is a sticky caustic in a freely falling flow and in this respect
it resembles the Bondi-Hoyle-Lyttleton (BHL) accretion (Bondi \& Hoyle 1944). 
In their case, accretion proceeds along a one-dimensional caustic 
called the accretion line. The accretion line is fed at each point by matter from 
an initially {\it plane-parallel} flow which is axisymmetric and has 
{\it zero} net angular momentum. 
In our case, the accretion disc is fed 
by an asymptotically {\it isotropic} inflow with a {\it non-zero} net angular 
momentum. In contrast to the BHL accretion line,
the radial momentum of matter
impinging the disc is directed inwards. 
As a result, we do not have the degenerate stagnation radius that exists 
in the BHL problem and leads to an infinite number of possible solutions. 

The fast disc is crucially different from the standard disc in many respects:

\begin{enumerate}

\item The fast disc forms in accretion flows with a net 
angular momentum $\bar{l}_z<2r_gc$.

\item The disc is small-scale, $\sim 10r_g$.

\item The disc is inviscid and accretes supersonically.

\item The energy release in the disc is caused by the infall absorption.

\item Most of the energy is released inside $3r_g$.

\item The disc is essentially relativistic; it does not admit any 
Newtonian approximation.

\item The disc has a low mass and its optical depth is not much 
above unity.

\item The disc dynamics depends on the angular momentum distribution of 
the feeding infall.

\end{enumerate}

\subsection{The generated spectrum}

As we argued in Section~1, small-scale discs are likely to form in 
wind-fed X-ray binaries and they are also possible in AGN. We now briefly 
discuss the expected X-ray spectrum emitted by the fast small-scale disc.
First we note that discs with $\dm>1$ ($L>\eta\LE$) have optical depths 
$\tT>1$. The emission then consists of two components: $L_h$ -- the hard
radiation from the top optically thin layer that radiates the vertical 
kinetic energy of the infall, and $L_s$ -- the radiation produced inside the 
optically thick disc where the horizontal velocities of the infall and the 
disc get equalised (see Section 2.3). Since $L_s$ comes out from the optically 
thick material, it is mostly soft radiation. 
By contrast, $L_h$ is generated by unsaturated Comptonization in the hot layer 
with the electron temperature $kT_e\sim 100$~keV (see eq.~12).
The Comptonized X-rays have  
a standard power-law spectrum with a break at $\sim 100$~keV.
The soft component $L_s$ is the source of seed photons for 
Comptonization in the hot layer.

The expected two-component spectrum is reminiscent of the observed spectra in 
galactic black hole sources and AGN (see Zdziarski 1999 for a review). 
The observed spectra are well fitted by two phenomenological models: 
(i) a hot accretion disc surrounded by a cold gaseous ring or 
(ii) a hot corona atop a cold disc.  
Both models are usually considered as modifications of the standard 
$\alpha$-disc in which the energy is released as a result of viscous stresses 
in a flow with high angular momentum (see Beloborodov 1999 for a review). 
The small-scale disc studied here generates a similar spectrum and it may 
provide a 
preferable explanation of the observed spectra in some cases. In general,
a Comptonised spectrum is degenerate,
i.e. it is insensitive to the heating mechanism and/or the accretion dynamics. 
Detailed studies of the spectrum features such as the
Fe K$\alpha$ line and the Compton reflection bump from the optically thick disc 
might help to break the degeneracy.

\subsection{Issues for further study}

\begin{enumerate}

\item Exact calculations of the emitted X-ray spectrum require detailed 
modelling of radiative transfer in the post-shock plasma.

\item At high $\dM$, the quasi-spherical infall gets so optically thick 
that it will trap the inner radiation and tend to advect it into the BH
(Begelman 1978). The trapping becomes efficient at $\dm\simgt 10$, when 
the time-scale of radiative diffusion through the infall is longer than 
the accretion time-scale. 
The corresponding reduction of the observed luminosity 
and the impact on the X-ray spectrum should be taken into account.
 
\item Throughout the paper, we assumed that the BH has a small spin. The similar 
problem can be solved for a rotating black hole.
 
\item We did not consider here the case $l>\lcr$ where 
accretion is stopped by the centrifugal barrier. At high $l$, the fast disc 
should transform into the standard accretion disc. This transition will be 
studied elsewhere (Beloborodov \& Illarionov, in preparation).

\end{enumerate}

%##########################################################################

\section*{Acknowledgments}

We thank J. Poutanen and R. Svensson for comments on the manuscript and 
A.G. Doroshkevich for discussions.
This work was supported by the Swedish Natural Science Research Council,
the Wenner-Gren Foundation for Scientific Research,
and RFBR grant 00-02-16135.

\appendix
\section{Conservation laws}

The barion conservation law reads $\nabla_k(\rho u^k)=0$ where $\nabla_i$
is the covariant derivative in Schwarzschild space-time, $\rho$ is 
matter density in its rest frame, $u^i=\dd x^i/\dd\lambda$ is matter
four-velocity, $\lambda$ is proper time, and $x^i$ are Schwarzschild 
coordinates. For the flow in the thin disc $u^\theta=0$, so that 
$u^i=(u^t,u^r,0,u^\varphi)$. The infall has 
4-velocity $\hat{u}^i$ and comoving density $\hat{\rho}$.

In coordinates $x^i$, the barion conservation law takes the form 
(e.g. Landau \& Lifshitz 1975) 
\begin{equation}
 \frac{1}{\sqrt{-g}}\frac{\partial}{\partial x^k}
 \left(\sqrt{-g}\rho u^k\right)=0,
\end{equation}
where $g$ is the determinant of the metric.
The non-zero components of Schwarzschild metric in the equatorial plane are
$g_{tt}=-(1-r_g/r)c^2$, $g_{rr}=r/(r-r_g)$, and
$g_{\theta\theta}=g_{\varphi\varphi}=r^2$. Then $\sqrt{-g}=r^2$.
Taking into account that the disc is steady, $\partial/\partial t=0$, and
axisymmetric, $\partial/\partial\varphi=0$, there are two terms in equation 
(A1), with $r-$ and $\theta-$derivatives. Integrating (A1) over $\theta$ 
between the two surfaces of the disc, one gets
\begin{equation}
  \frac{\dd\dM}{\dd r}=-4\pi r^2\hat{\rho}\hat{u}^\theta, \quad
  \dM\equiv 2\pi r \Sigma u^r.
\end{equation}
Here $\Sigma=\rho\sqrt{g_{\theta\theta}}\Delta\theta$ is the surface density 
of the disc (measured in its rest frame), $\Delta\theta=2H/r$ is the angular
thickness of the disc.

The energy and momentum conservation reads $\nabla_iT^i_k=0$ where $T^i_k$ 
is the stress-energy tensor. In coordinates $x^i$ this 
equation reads (Landau \& Lifshitz 1975)
\begin{equation}
  \frac{1}{\sqrt{-g}}\frac{\partial }{\partial x^k}\left(\sqrt{-g}T^k_i\right) 
  =\frac{1}{2}\frac{\partial g_{kl}}{\partial x^i} T^{kl}.
\end{equation}
The disc internal energy and pressure are small (heat is radiated away)
and we neglect their contribution to $T^i_k$. We also neglect viscous stresses.
$T^i_k$ then takes the simple form 
\beq
\nonumber
T^i_k=\rho c^2u^iu_k.
\eeq
Analogously, the stress-energy tensor of the infall through the disc surface 
is $T^i_k=\hat\rho c^2\hat{u}^i\hat{u}_k$. The stress-energy tensor associated
with the vertical radiation flux from the disc is $T^i_k=u^iq_k+u_kq^i$ where
$q^i$ is the four-flux vector (e.g. Misner et al. 1973). Performing vertical 
integration across the disc and using equation (A2), we get
\begin{equation}
  \frac{\dd}{\dd r}\left(\dM u_i\right)-\hat{u}_i\frac{\dd\dM}{\dd r}
  -\frac{2\pi F}{c^2}ru_i
  =\frac{1}{2}\frac{\partial g_{kl}}{\partial x^i} u^ku^l\frac{\dM}{u^r}.
\end{equation}
Here $F=-(\dd L/\dd r)(c^2/u_t2\pi r)$ is the radiation flux from the two faces 
of the disc (measured in the rest frame of the accreting matter).

For $x^i=t$ and $x^i=\varphi$ equation (A4) states the conservation of 
energy, $E=-u_t$, and angular momentum, $j=u_\varphi$, respectively, 
and in both cases the right-hand-side vanishes. The radiative term on the
left hand-side is $\sim L/\dM c^2\ll 1$ smaller compared to the other two 
terms and it can be neglected in the conservation law of angular momentum. 

For $x^i=r$, the right-hand-side in eqiation~(A4) is non-zero; simple
algebra (using $u_tu^t+u_ru^r+u_\varphi u^\varphi=-c^2$) gives
\beq
\nonumber
  \frac{\partial g_{kl}}{\partial r}u^ku^l
  =-\frac{r_gc^2}{r(r-r_g)}-\frac{2r_g}{r^2}u_r^2
   +\frac{(2r-3r_g)}{r^3(r-r_g)}u_\varphi^2.
\eeq
Substituting this relation into (A4) and neglecting the radiative term, 
we get equation (21).

%##############################################################################

\section{The luminosity capture by the black hole}

Assume that each element of the disc $\dd S$
emits radiation isotropically in the rest frame of the accreting matter. 
Let $L_c$ be the luminosity
of $\dd S$ measured in the rest frame, $L_{\rm loc}$ -- the luminosity 
of $\dd S$ measured by the local stationary observer (LSO), ${\bbeta}$ --
the local disc velocity measured by the LSO (in units of $c$), and 
$\gamma=(1-\beta^2)^{-1/2}$. For a steady disc we have the relation (see 
e.g. Rybicki \& Lightman 1979, chapter 4.8)
\be
  \frac{\dd L_{\rm loc}}{\dd\Omega}
=\frac{L_c}{4\pi\gamma^4(1-\bbeta\cdot{\bf\Omega})^3},
\ee
where ${\bf\Omega}$ is the unit vector specifying direction of an emitted
photon in the LSO frame and $\dd\Omega$ is the element of the solid angle. 
The local luminosity emitted in $4\pi$,
as measured by LSO, is equal to the rest frame luminosity, $L_{\rm loc}=L_c$. 

Consider the local tetrade 
$(e_{(t)}^i,e_{(r)}^i,e_{(\theta)}^i,e_{(\varphi)}^i)$ and let 
$(\beta^{(r)},0,\beta^{(\varphi)})$ and  
$(\Omega^{(r)},\Omega^{(\theta)},\Omega^{(\varphi)})$ be the tetrade components
of $\bbeta$ and ${\bf\Omega}$, respectively. The $\beta^{(\alpha)}$ 
($\alpha=1,2,3$) is related to the four-velocity of the disc, $u^i$, by 
$\beta^{(\alpha)}=(u^\alpha/u^t)(-g_{\alpha\alpha}/g_{tt})^{1/2}$, and
$\gamma=u^t(-g_{tt})^{1/2}$. The $\Omega^{(\alpha)}$ can be written as
$\Omega^{(r)}=-\cos\hat\theta$, 
$\Omega^{(\theta)}=-\sin\hat\theta\sin\hat\varphi$,
$\Omega^{(\varphi)}=\sin\hat\theta\cos\hat\varphi$, and
$\dd\Omega=\sin\hat\theta\dd\hat\theta\dd\hat\varphi$. 
Equation (27) now takes the form
\beq
\nonumber
 \frac{\dd L_{\rm loc}}{\dd\hat\theta\dd\hat\varphi}
=\frac{L_{\rm loc}\sin\hat\theta}{4\pi\gamma^4[1+\beta^{(r)}\cos\hat\theta
-\beta^{(\varphi)}\sin\hat\theta\cos\hat\varphi]^3}.
\eeq
The black hole absorbs radiation emitted within a cone 
$\hat\theta<\hat\theta_{\rm abs}(r)$ (see Misner et al. 1973), where
\beq
\nonumber
  \sin^2\hat\theta_{\rm abs}=\frac{27r_g^2}{4r^2}
  \left(1-\frac{r_g}{r}\right).
\eeq
$\hat\theta_{\rm abs}<\pi/2$ at $r>(3/2)r_g$ and $\hat\theta_{\rm abs}>\pi/2$ 
at $r<(3/2)r_g$. At $r=r_g$, $\hat\theta_{\rm abs}=\pi$, i.e. all 
radiation is absorbed.

The captured fraction of radiation emitted at a radius $r$ is 
\beq
\kappa(r)=\int_0^{2\pi}\int_0^{\hat\theta_{\rm abs}}
\frac{\sin\hat\theta\,\dd\hat\theta\,\dd\hat\varphi}
{4\pi\gamma^4[1+\beta^{(r)}\cos\hat\theta
-\beta^{(\varphi)}\sin\hat\theta\cos\hat\varphi]^3}.
\eeq
Here $\beta^{(r)}=u^r/E$, $\beta^{(\varphi)}=(cj/rE)(1-r_g/r)^{1/2}$, and
$\gamma=(E/c^2)(1-r_g/r)^{-1/2}$.
The released and captured power from $\dd S$ as measured at infinity is 
$L_{\rm loc}(-g_{tt})$ and $\kappa L_{\rm loc}(-g_{tt})$, 
respectively. The resulting radial distribution of the observed luminosity is 
\beq
 \frac{\dd L_{\rm obs}}{\dd r}=\frac{\dd L}{\dd r}\left[1-\kappa(r)\right].
\eeq

Equation (B3) is exact for an optically thin disc only. An optically thick
disc partly intercepts its own radiation owing to the gravitational bending
of light near the black hole, and reemits this radiation. As a result of this
self-illumination, the radial distribution of the disc luminosity changes.
We here neglect these effects and use equations (B2) and (B3) to estimate the 
captured luminosity.

\end{document}